\documentclass[sigconf]{acmart}
\AtBeginDocument{%
  \providecommand\BibTeX{{%
    \normalfont B\kern-0.5em{\scshape i\kern-0.25em b}\kern-0.8em\TeX}}}

\copyrightyear{2023} 
\acmYear{2023} 
\setcopyright{acmlicensed}
\acmConference[SIGIR '23]{Proceedings of the 46th International ACM SIGIR Conference on Research and Development in Information Retrieval}{July 23-27, 2023}{Taipei, Taiwan, China}
\acmBooktitle{Proceedings of the ACM SIGIR Conference 2023 (SIGIR '23), July 23-27, 2023, Taipei, Taiwan, China}
\acmPrice{15.00}
\acmDOI{XXXXXXXXXXXXX}
\acmISBN{XXXXXXXXXXXXX}

\usepackage{xspace}
\usepackage{amsthm}
\usepackage{amsmath}
\usepackage{enumitem}
\usepackage{booktabs}
\usepackage{multirow}
\usepackage{graphicx}
\usepackage[normalem]{ulem}
\useunder{\uline}{\ul}{}
\usepackage{float}
\usepackage{caption}
\usepackage{subcaption}
\usepackage{balance}
\usepackage[toc,page]{appendix}

\newcommand{\modelname}{\textsf{GraphDA}\xspace}
\newcommand{\subvariant}{\textsf{Enhanced-UI}\xspace}

\usepackage{amsmath}
\DeclareMathOperator*{\argmax}{arg\,max}

\begin{document}

\title{Graph Collaborative Signals Denoising and Augmentation for Recommendation}

\author{Ziwei Fan, Ke Xu}
\affiliation{%
  \institution{University of Illinois Chicago}
  \country{USA}
}
\email{{zfan20, kxu25}@uic.edu}

\author{Zhang Dong}
\affiliation{%
  \institution{Amazon Ads}
  \country{USA}
}
\email{andydong@amazon.com}

\author{Hao Peng}
\orcid{0000-0001-7422-630X}
\authornote{Corresponding author}
\affiliation{%
  \institution{Beihang University}
  \country{China}
}
\email{penghao@act.buaa.edu.cn}

\author{Jiawei Zhang}
\affiliation{%
  \institution{University of California, Davis}
  \country{USA}
}
\email{jiawei@ifmlab.org}

\author{Philip S. Yu}
\affiliation{%
  \institution{University of Illinois Chicago}
  \country{USA}
}
\email{psyu@uic.edu}

\renewcommand{\shortauthors}{Fan, et al.}

\begin{abstract}
  Graph collaborative filtering (GCF) is a popular technique for capturing high-order collaborative signals in recommendation systems. However, GCF's bipartite adjacency matrix, which defines the neighbors being aggregated based on user-item interactions, can be noisy for users/items with abundant interactions and insufficient for users/items with scarce interactions. Additionally, the adjacency matrix ignores user-user and item-item correlations, which can limit the scope of beneficial neighbors being aggregated. 
  
  In this work, we propose a new graph adjacency matrix that incorporates user-user and item-item correlations, as well as a properly designed user-item interaction matrix that balances the number of interactions across all users. To achieve this, we pre-train a graph-based recommendation method to obtain users/items embeddings, and then enhance the user-item interaction matrix via top-K sampling. We also augment the symmetric user-user and item-item correlation components to the adjacency matrix. Our experiments demonstrate that the enhanced user-item interaction matrix with improved neighbors and lower density leads to significant benefits in graph-based recommendation. Moreover, we show that the inclusion of user-user and item-item correlations can improve recommendations for users with both abundant and insufficient interactions.
  The code is in \url{https://github.com/zfan20/GraphDA}. 
\end{abstract}




\maketitle

\begin{figure}
    \centering
    \begin{subfigure}[b]{0.22\textwidth}
         \centering
         \includegraphics[width=1\textwidth]{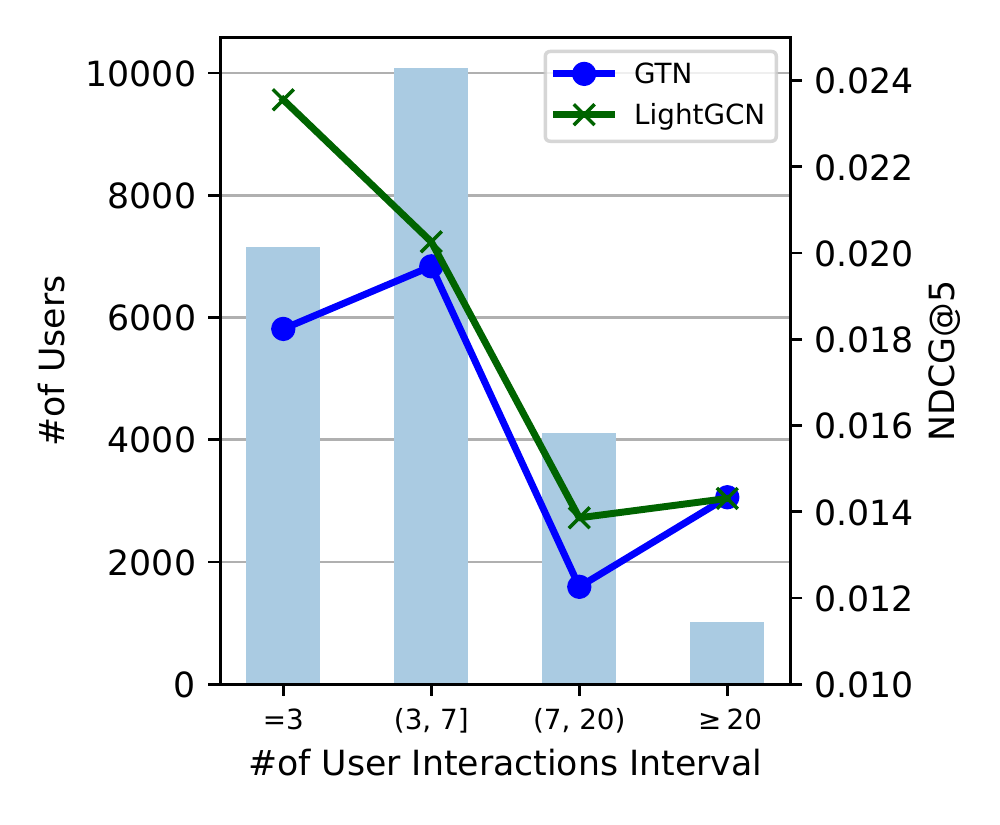}
         \caption{Users}
         \label{fig:beauty_users_gcf}
     \end{subfigure}\hfill
     \begin{subfigure}[b]{0.22\textwidth}
         \centering
         \includegraphics[width=1\textwidth]{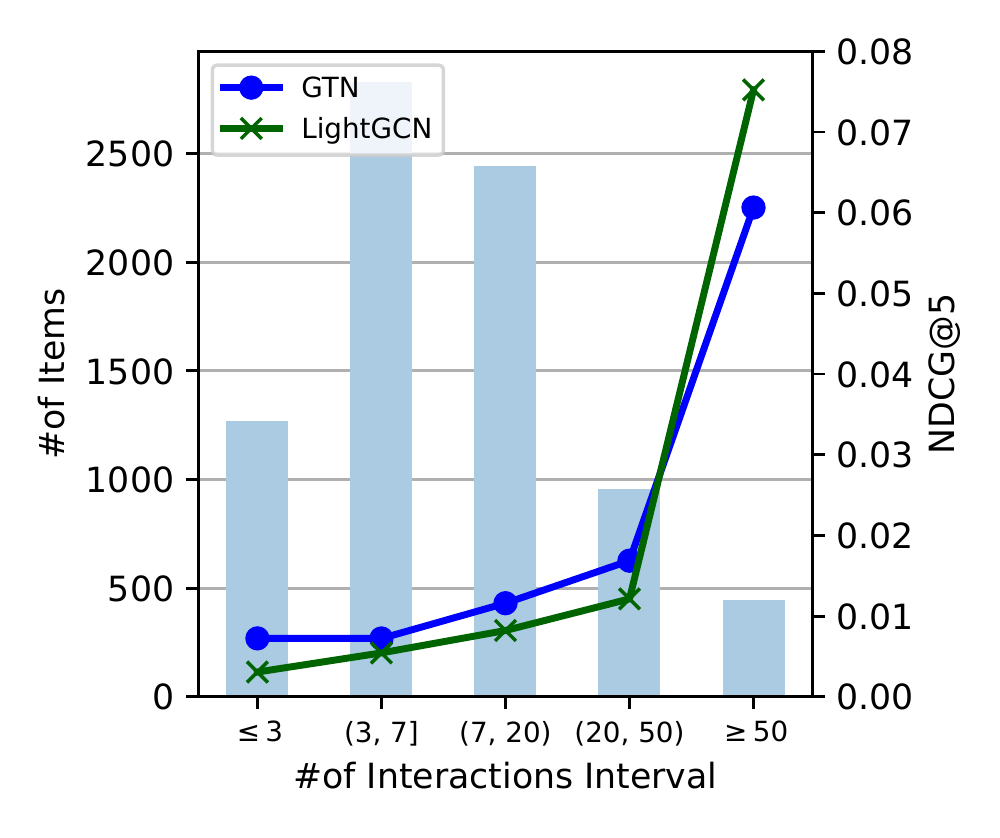}
         \caption{Items}
         \label{fig:beauty_items_gcf}
     \end{subfigure}
    \caption{The interactions amount distribution of users\slash items  (bar) and corresponding NCDG@5 (line) on Amazon Beauty dataset by two graph models LightGCN~\cite{he2020lightgcn} and GTN~\cite{fan2022graph}.}
    \label{fig:intro_lightgcn_perf_seqlen_dist}
\end{figure}
\begin{figure}
    \centering
    \includegraphics[width=0.9\columnwidth]{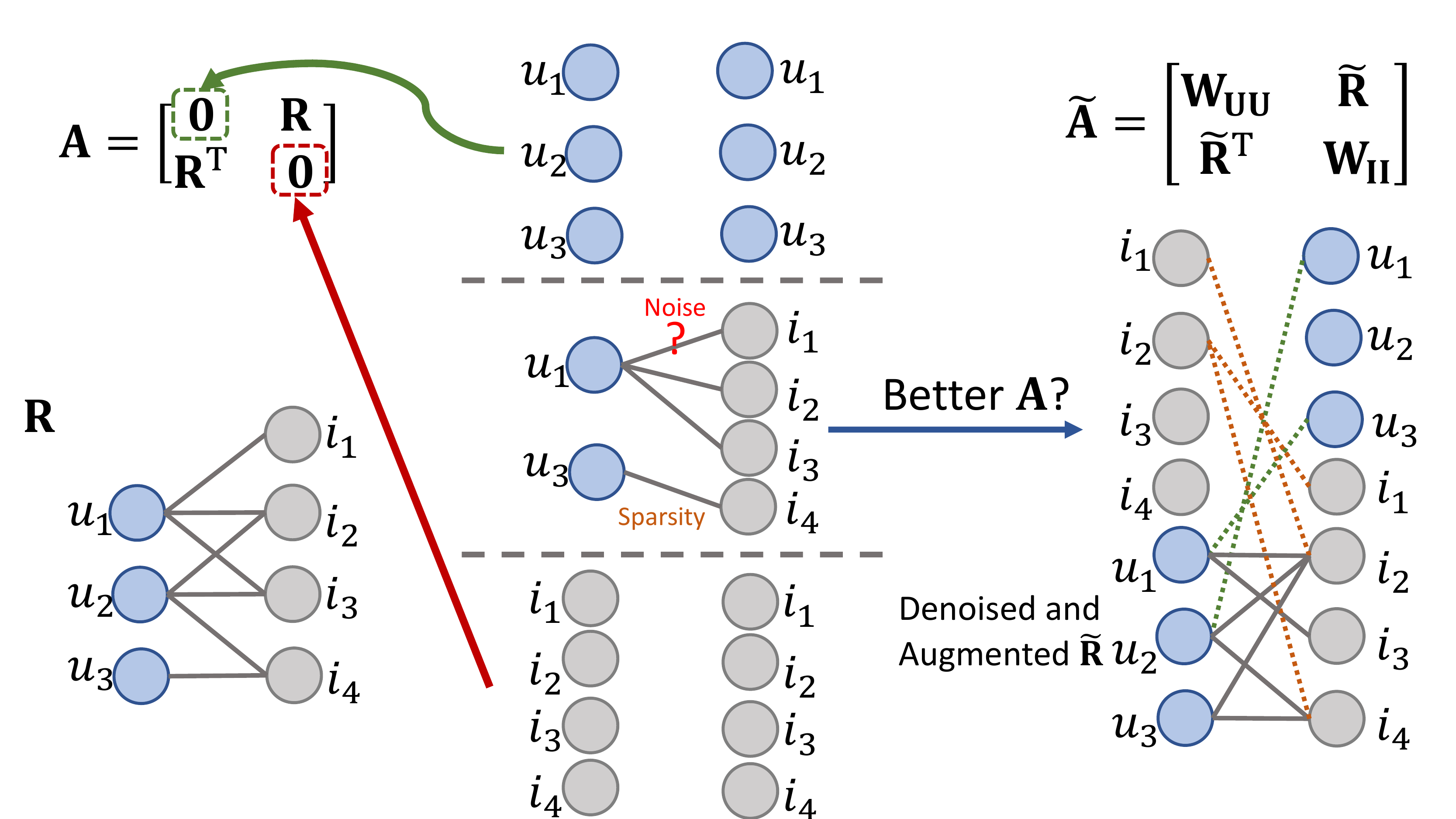}
    \caption{Motivation figure with a toy example. The $\mathbf{A}$ and $\mathbf{R}$ are the bipartite adjacency matrix and the user-item interaction matrix for graph-based recommendations, respectively.}
    \label{fig:movitation}
\end{figure}

\section{Introduction}
Graph-based recommender systems use neighborhood information to infer user/item embeddings, where the adjacency matrix defines the neighborhood structure. High-order collaborative signals are typically aggregated by stacking multiple layers~\cite{he2020lightgcn, wang2019neural, liu2021federated, lin2020fill, pinsage2018ying, berg2017graph, chen2019personalized, lin2022phish, yang2022large}. However, the quality of the neighborhood information depends on the definition of the adjacency matrix. The widely adopted adjacency matrix is built upon the user-item interaction matrix, which potentially encounters noises~\cite{fan2022graph, wu2021self, 10.1145/3459637.3482092}, sparsity~\cite{bayer2017generic, he2016ups, hao2021pre, liu2020basket}, biases~\cite{zhu2020measuring, abdollahpouri2019unfairness, chen2020bias}, and long tail~\cite{park2008long, clauset2009power, milojevic2010power} issues. As shown in Fig.~(\ref{fig:intro_lightgcn_perf_seqlen_dist}), we can observe that both users and items are following the long-tail distribution, where majority of users\slash items have limited interactions. Moreover, one counter-intuitive observation in Fig.~(\ref{fig:intro_lightgcn_perf_seqlen_dist}) is that users with rich interactions (i.e., active users) are poorly modeled, compared with users with scarce interactions (i.e., inactive users)~\cite{lam2008addressing}. Arguably, the underlying reason is that highly active users have abundant noisy interactions, which even might be harmful to the user preference modeling. Furthermore, more noises are introduced when the graph model stacks more layers of graph convolutions~\cite{fan2022graph}. From the item side, we can observe that items with limited interactions are performed unsatisfactorily.

Based on these observations, we argue that the current definition of the bipartite adjacency matrix in graph-based recommender systems is inadequate. As shown in Figure~(\ref{fig:movitation}), the bipartite adjacency matrix $\mathbf{A}$ is constructed directly from the user-item interaction matrix $\mathbf{R}$ to define the neighborhood structure for users/items. However, $\mathbf{R}$ suffers from noisy and sparse interactions, making it insufficient to represent inactive users/items. Additionally, the bipartite adjacency matrix $\mathbf{A}$ overlooks the user-user~\cite{zhu2019improving, meng2021graph} and item-item correlations~\cite{christakopoulou2016local, hu2019hybrid, ning2011slim} in the neighborhood definition, even in the enhanced solution~\cite{yang2021enhanced, chen2021structured}. Although high-order collaborative signals can uncover these correlations via multi-hop message passing, recent studies have shown that long-distance message passing can create new learning problems and lead to suboptimal representations~\cite{chen2020measuring}. Therefore, we propose a novel adjacency matrix design to improve the graph-based recommendation.

It is challenging to properly design a better adjacency matrix. Some relevant distillation methods~\cite{zheng2021cold, jin2021graph} learn a smaller but useful graph data for graph modeling. However, one significant difference with distillation methods is that the GCF utilizes users and items IDs as inputs, and thus existing works of graph condensation are not applicable in our GCF setting. Moreover, these distillation methods assume the availability of features while users' features are sometimes not accessible due to privacy constraints.

To this end, we propose a pre-training and enhancing pipeline framework, namely \modelname, to denoise and augment the user-item matrix. Within \modelname, we capture the user-user and item-item correlations in the bipartite adjacency matrix for the GCF. Specifically, we first pre-train an encoder to generate the users\slash items embeddings from existing user-item interactions. With pre-trained embeddings, we adopt the top-K sampling process to generate the denoised and augmented user-item matrix, non-zero user-user and item-item correlations. Our contributions include:
\begin{itemize}[leftmargin=*]
    \item We investigate the deficiency of the existing definition of the bipartite adjacency matrix for GCF and study the potential of introducing a better adjacency matrix.
    \item We propose a better adjacency matrix generation for the graph-based recommendation, with a novel pipeline \modelname for denoising for active users and augmenting for inactive users.
    \item Comprehensive experiments show that the proposed \modelname significantly benefits the graph-based recommendation, especially on highly active users and inactive users, who demand denoising and augmentation, respectively. 
\end{itemize}

\section{Graph Collaborative Filtering}
In the graph collaborative filtering~(GCF), we denote the user set as $\mathcal{U}$ and the item set as $\mathcal{I}$, where the  user and item are indexed by $u$ and $i$. With either implicit or explicit feedback, the user-item interaction matrix is given as $\mathbf{R}\in \mathbb{R}^{|\mathcal{U}|\times|\mathcal{I}|}$, where $\mathbf{R}_{ui}$ denotes the feedback on the item $i$ given by the user $u$. For example, $\mathbf{R}_{ui}$ of implicit feedback is either 1 or 0. As the $\mathbf{R}$ is a user-item bipartite graph, the adjacency matrix is further formatted as:
    $\mathbf{A} =  \left[ \begin{array}{cc}
\mathbf{0} & \mathbf{R} \\ \mathbf{R}^\top & \mathbf{0} \end{array} \right],$
where $\mathbf{A}\in\mathbb{R}^{(|\mathcal{U}|+|\mathcal{I}|)\times(|\mathcal{I}|+|\mathcal{U}|)}$. 
This problem can also be interpreted as the link prediction problem between user and item nodes.
The user and item embeddings are randomly initialized and optimized with existing user-item interactions. Specifically, we denote the user and item embeddings table as $\mathbf{E}\in \mathbb{R}^{(|\mathcal{U}|+|\mathcal{I}|)\times d}$, where $d$ denotes the latent dimension of embeddings. GCF incorporates high-order collaborative signals~\cite{wang2019neural, he2020lightgcn} by stacking multiple layers of graph convolution on the user-item adjacency matrix $\mathbf{A}$.
Specifically, the output embeddings generation process with $N$ graph convolution layers is given as:
\begin{align}
\label{eq:emb_gen}
    \mathbf{E}^{(N)} = \text{Encoder}(\mathbf{A}, \mathbf{E}) = (\mathbf{L})^{N-1}\mathbf{E}^{(0)},
\end{align}
where $\mathbf{E}^{(0)}=\mathbf{E}$, $\mathbf{L}$ refers to the bipartite Laplacian matrix, which is defined as the normalized symmetric Laplacian matrix $\mathbf{D}^{-\frac{1}{2}}\mathbf{A}\mathbf{D}^{-\frac{1}{2}}$, and $\mathbf{D}$ is the degree matrix. The representative work LightGCN~\cite{he2020lightgcn} averages the generated embeddings over all layers as the final output embedding. The user preference prediction between the user $u$ and item $i$ is given as:
\begin{align}
    P(i|u, \mathbf{A}) = \sigma \left(\mathbf{e}_u^\top \mathbf{e}_i\right) \ \ \text{where }\ i\in\mathcal{I}\setminus\mathcal{I}_u^+,
\end{align}
where $\sigma(\cdot)$ denotes the sigmoid activation function, $\mathcal{I}_u^+$ denotes the observed interacted item set by the user $u$, $\mathbf{e}_u$ and $\mathbf{e}_i$ are the user and item output embeddings of $\mathbf{E}^{(N)}$.


\begin{figure}
    \centering
    \includegraphics[width=1\columnwidth]{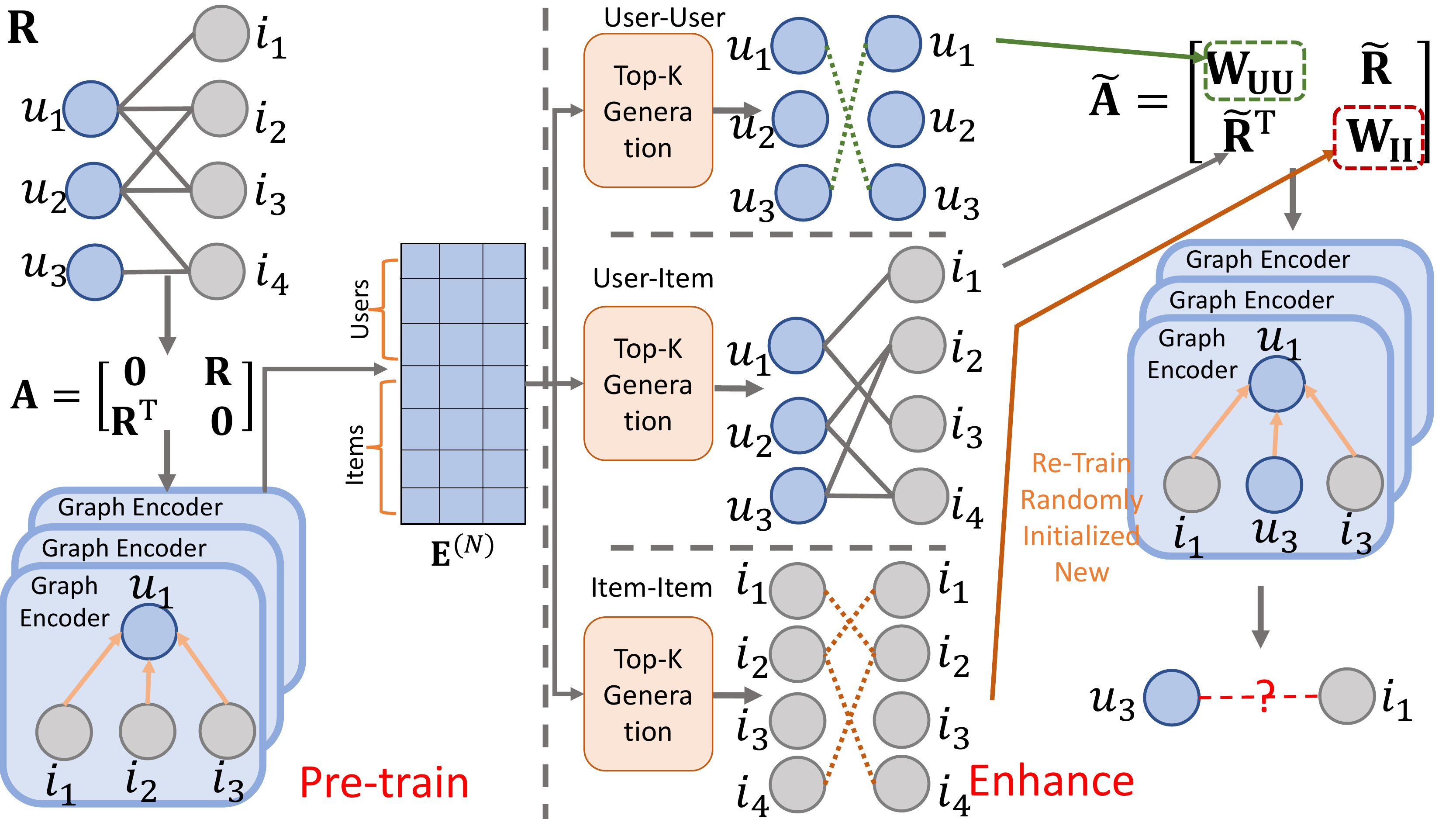}
    \caption{Workflow Diagram of \modelname. \modelname consists of two steps: 1. the pre-train step infers users\slash items embeddings; 2. utilize embeddings to generate top-K neighbors for the user-item component, the user-user component, and the item-item component, the enhanced adjacency matrix is used to re-train a graph encoder.}
    \label{fig:workflow}
\end{figure}
\section{Proposed Framework}
In this section, we introduce our proposed framework \modelname with both user-item interaction matrix, user-user and item-item correlations enhancements. The framework consists of two steps, including the users\slash items representations pre-trained from a graph encoder and neighbors generation processes for enhancement. 

\subsection{Pre-Trained Users\slash Items Embeddings}
With the arguably imperfect graph Laplacian matrix $\mathbf{L}$ from the original adjacency matrix $\mathbf{A}$, we pre-train a graph encoder to obtain the users\slash items representations, which is shown as pre-train in the left part of Fig.~(\ref{fig:workflow}). Specifically, we use $N$ graph convolution layers to obtain users\slash items embeddings $\mathbf{E}^{(N)}$ as described in Eq.~(\ref{eq:emb_gen}). The pre-train step is optimized with the training data using the BPR loss as:
\begin{equation}
    \mathcal{L} = -\sum_{(u,i^+,i^-\in \mathbf{R})}\log \sigma(\mathbf{e}_u^\top \mathbf{e}_{i^+}-\mathbf{e}_u^\top \mathbf{e}_{i^-}),
\end{equation}
where $i^+$ denotes the positively interacted item of the user $u$, $i^{-}$ is a sampled negative item without interaction with the user $u$, and $\mathbf{e}_u$ and $\mathbf{e}_i$ are the user and item output embeddings in $\mathbf{E}^{(N)}$. Note that several encoders with various architectures modeling user-item interactions can be the alternative choice, such as the classical matrix factorization~\cite{10.5555/1795114.1795167}. 

\subsection{Enhanced Bipartite Adjacency}
\label{sec:enhanced_adj}
The pre-trained $\mathbf{E}^{(N)}$ encodes user-item collaborative signals. However, its crucial component in GCF, i.e., the adjacency matrix $\mathbf{A}$, is arguably less satisfactory for users\slash items embeddings learning, due to the biased interactions observed as the long-tailed distribution, noisy interactions for active users, and the ignoring of direct user-user and item-item correlations. Specifically, we enhance the adjacency matrix in three components, as shown in the central component of Fig.~(\ref{fig:workflow}).

\noindent {\bf User-item Interactions Enhanced.} With the pre-trained $\mathbf{E}^{(K)}$, we generate the top-K neighbors for both users and items. For users, the top-K neighbors define the preference while neighbors for items represent the concept of the user group being marketed. From the user side, we define a hyper-parameter $U_k$ to control the number of neighbors being selected. For all users, the number of neighbors is the same $U_k$. With equal number of neighbors, users with abundant neighbors in the original data are denoised while users with scarce neighbors are augmented. Specifically, for the user $u$, the top-K neighbors are generated by selecting the top $U_k$ elements with largest values from the output scores $\mathbf{e}_u^\top \mathbf{E}_{\mathcal{I}}^{(N)}$, i.e.,:
\begin{equation}
    \argmax_{\{i_1,i_2,\dots,i_{U_k}\in\mathcal{I}\}} \mathbf{e}_u^\top \mathbf{E}^{(N)}_{\mathcal{I}},
\end{equation}
where $\mathbf{E}^{(N)}_{\mathcal{I}}$ denotes output item embeddings. Similarly, for the item side, we also define the hyper-parameter $I_k$ to generate the top-K neighbors with the similar process. 
We adopt the union of generated user-item interactions from both user and item sides to obtain the enhanced $\widetilde{\mathbf{R}}$. 

\noindent {\bf User-User and Item-Item Correlations.} For the recommender system dataset with unknown user-use/item-item interactions, the corresponding sub-matrices in conventional adjacency matrix $\mathbf{A}$ representation (as indicated at the beginning of Section 2) will be filled with zeros. We propose to complement these two all-zero sub-matrices with $\mathbf{W}_{UU}$ and $\mathbf{W}_{II}$ to further enhance the adjacency matrix. Specifically, for a user $u$, we extract the top-$UU_k$ similar users as:
\begin{equation}
     \argmax_{\{u_1,u_2,\dots,u_{UU_k}\in\mathcal{U}\}} \mathbf{e}_u^\top \mathbf{E}^{(N)}_{\mathcal{U}},
\end{equation}
where $\mathbf{E}^{(N)}_{\mathcal{U}}$ denotes the output user embeddings, and $\mathbf{W}_{UU}[u, u_k]=1\ \ \text{for } u_k\in\{u_1,u_2,\dots,u_{UU_k}\}$. The similar process is conducted for $\mathbf{W}_{II}$ with $II_k$ to control the number of similar items being chosen. Note that we enforce the $\mathbf{W}_{UU}$ and $\mathbf{W}_{II}$ to be symmetric. 

\noindent {\bf GCF Re-learn with Enhanced Bipartite Adjacency.} With the enhanced graph adjacency matrix, we re-learn the graph encoder~(i.e., randomly initialized) to generate embeddings for user-item interaction predictions, which is shown in the right part in Fig.~(\ref{fig:workflow}). To better illustrate each component's contribution, we propose two version of enhancements, \subvariant and \modelname. The \subvariant only adopts the user-item enhanced interactions, i.e., its adjacency matrix $\widetilde{\mathbf{A}}=\left[ \begin{array}{cc}
\mathbf{0} & \widetilde{\mathbf{R}} \\ \widetilde{\mathbf{R}}^\top & \mathbf{0} \end{array} \right]$. The complete version \modelname includes user-user and item-item correlations, having the adjacency matrix $\widetilde{\mathbf{A}}=\left[ \begin{array}{cc}
\mathbf{W}_{UU} & \widetilde{\mathbf{R}} \\ \widetilde{\mathbf{R}}^\top & \mathbf{W}_{II} \end{array} \right]$. 

\section{Experiments}
This section presents experiments for demonstrating the effectiveness of the proposed framework \modelname. We answer following research questions~(RQs): {\bf RQ1: }Does \modelname achieve better recommendation performances than existing baselines? {\bf RQ2: }What are the effects of hyper-parameters and each component in \modelname? {\bf RQ3: }Does \modelname achieve the goals of denoising and augmentation simultaneously?


\subsection{Experimental Settings}
\noindent \textbf{Datasets.} We use the public Amazon Reviews dataset~\cite{mcauley2015image} with three benchmark categories~\cite{fan2022graph, he2020lightgcn, wang2019neural, li2020time}, including: (1) \textit{Beauty} has 22,363 users, 12,101 items, and 198,502 interactions with 0.05\% density; (2) \textit{Toys and Games}~(Toys) has 19,412 users, 11,924 items, and 167,597 interactions with 0.07\% density; (3) \textit{Tools and Home}~(Tools) has 16,638 users, 10,217 items, and 134,476 interactions with 0.08\% density. We follow the 5-core setting as existing works on users and the same transformation~\cite{fan2022graph, he2020lightgcn, wang2019neural} of treating the existence of reviews as positives. We sort each user's interactions chronologically and adopt the leave-one-out setting, with the last interacted item for testing and the second last interaction for validation.

\begin{table*}[h]
\centering
\caption{Overall Comparison Table in HR@20 and NDCG@20. The best baseline and best model are underlined and in bold. `Improv.' indicates the relative improvements over the best baseline. }
\label{tab:overall_performances}
\resizebox{\textwidth}{!}{%
\begin{tabular}{c|cccc|cccc|cccc|cccc}
\hline
Dataset & \multicolumn{4}{c|}{Beauty} & \multicolumn{4}{c|}{Toys} & \multicolumn{4}{c|}{Tools} & \multicolumn{4}{c}{Office} \\ \hline
Metric & H@10 & N@10 & H@20 & N@20 & H@10 & N@10 & H@20 & N@20 & H@10 & N@10 & H@20 & N@20 & H@10 & N@10 & H@20 & N@20 \\ \hline
NGCF & 0.0447 & 0.0232 & 0.0724 & 0.0299 & 0.0461 & 0.0251 & 0.0672 & 0.0306 & 0.0329 & 0.0179 & 0.0480 & 0.0216 & 0.0261 & 0.0159 & 0.0453 & 0.0208 \\
UltraGCN & 0.0451 & 0.0234 & 0.0728 & 0.0304 & 0.0464 & 0.0250 & 0.0675 & 0.0308 & 0.0331 & 0.0179 & 0.0481 & 0.0217 & 0.0302 & 0.0171 & 0.0471 & 0.0210 \\
GTN & 0.0446 & 0.0230 & 0.0680 & 0.0289 & 0.0453 & 0.0248 & 0.0661 & 0.0301 & {\ul 0.0337} & {\ul 0.0184} & {\ul 0.0484} & {\ul 0.0221} & 0.0283 & 0.0161 & 0.0453 & 0.0204 \\
LightGCN & {\ul 0.0471} & {\ul 0.0244} & {\ul 0.0730} & {\ul 0.0309} & {\ul 0.0512} & {\ul 0.0273} & {\ul 0.0716} & {\ul 0.0325} & 0.0334 & 0.0182 & 0.0482 & 0.0219 & {\ul 0.0355} & {\ul 0.0197} & {\ul 0.0522} & {\ul 0.0238} \\ \hline
\subvariant & 0.0486 & 0.0252 & 0.0755 & 0.0317 & 0.0530 & 0.0276 & 0.0765 & 0.0335 & 0.0364 & 0.0195 & 0.0527 & 0.0236 & 0.0363 & 0.0208 & \textbf{0.0565} & 0.0259 \\
Improv. & +3.2\% & +3.3\% & +3.1\% & +2.9\% & +3.5\% & +1.1\% & +6.8\% & +3.3\% & +8.0\% & +6.0\% & +4.3\% & +6.6\% & +2.3\% & +5.6\% & +8.2\% & +8.8\% \\ \hline
\modelname & \textbf{0.0514} & \textbf{0.0264} & \textbf{0.0804} & \textbf{0.0336} & \textbf{0.0549} & \textbf{0.0289} & \textbf{0.0795} & \textbf{0.0347} & \textbf{0.0373} & \textbf{0.0205} & \textbf{0.0532} & \textbf{0.0245} & \textbf{0.0383} & \textbf{0.0225} & 0.0561 & \textbf{0.0270} \\
Improv. & +9.1\% & +8.2\% & +8.7\% & +7.9\% & +7.2\% & +5.9\% & +11.1\% & +6.9\% & +10.7\% & +11.4\% & +5.4\% & +10.8\% & +7.9\% & +14.2\% & +7.5\% & +13.4\% \\ \hline
\end{tabular}%
}
\end{table*}

\noindent \textbf{Evaluations.} We adopt the widely used standard ranking evaluation metrics to evaluate the averaged ranking performance over all users, including Recall@N and NDCG@N, which are widely used in existing works~\cite{fan2022graph, he2020lightgcn, wang2019neural}. For the fair comparison without the sampling bias, we adopt the \textbf{all items ranking evaluation}~\cite{krichene2020sampled}. 
Moreover, we use \textbf{only one training negative item} during the training process. 
Recall@N measures the correctness of ranking the ground-truth items in the top-N list, regardless of the ranking position. NDCG@N extends to further give higher weights to higher ranking positions. 
We only report the $N=20$ due to limited space.

\noindent \textbf{Baselines.} We compare several graph-based recommendation methods, including LightGCN~\cite{he2020lightgcn}, NGCF~\cite{wang2019neural}, UltraGCN~\cite{mao2021ultragcn}, and GTN~\cite{fan2022graph}. 

\noindent \textbf{Hyper-Parameters Grid Search.} For all methods, we use the dimension as 128 and learning rate as $0.001$. We search the L2 regularization weight in $\{0.0, 0.01, 0.001, 0.0001\}$. For NGCF, we search the node and message dropouts from $\{0.1, 0.3, 0.5\}$. For GTN, we search its specific $\beta$ and $\lambda$ from $\{0.001, 0.005, 0.01, 0.05, 0.1\}$. For UltraGCN, we search their weights from $\{1, 1e-3, 1e-5, 1e-7\}$ and negative weights from $\{10, 50, 100\}$. All hyper-parameters are grid-searched and the test performance is reported based on the best validation performance. We search the $U_K$ and $I_K$ from $\{0, 3, 5, 7, 9\}$. We also search the $UU_K$ and $II_K$ from $\{0, 3, 5, 7\}$. 


\subsection{Overall Comparison~(RQ1)}
We obtain several observations from the overall comparison Table~\ref{tab:overall_performances}:
\begin{itemize}[leftmargin=*]
    \item \textbf{Enhancing the bipartite Laplancian matrix benefits the graph-based recommendation.} The proposed framework \modelname and its variant \subvariant achieve significant improvements over existing graph-based recommendation methods. The superiority of \modelname and \subvariant demonstrate the benefits of enhancing the bipartite Laplancian matrix. 
    \item \textbf{User-user and item-item correlations are beneficial.} The user-user and item-item correlations enhancements on the bipartite Laplacian contributes to further performance improvements. When we compare \subvariant~(with only the user-item component) and \modelname~(with additional user-user and item-item correlations), \modelname outperforms \subvariant.
    \item Among existing graph recommendation methods, LightGCN achieves the best baseline performance. The second best baseline is GTN or UltraGCN, depending on datasets. Due to its simplicity, LightGCN can achieve satisfactory performances, and GTN proposes the trend filtering mechanism to denoise the user interactions in the LightGCN framework.  UltraGCN infuses the item-item similarities in the graph recommendation. These comparisons demonstrate the necessity of incorporating item-item correlations and user interactions denoising process in the graph recommendation.
\end{itemize}

\subsection{Hyper-Parameters Sensitivity~(RQ2)}
\begin{figure}
    \centering
    \begin{subfigure}[b]{0.2\textwidth}
         \centering
         \includegraphics[width=1\textwidth]{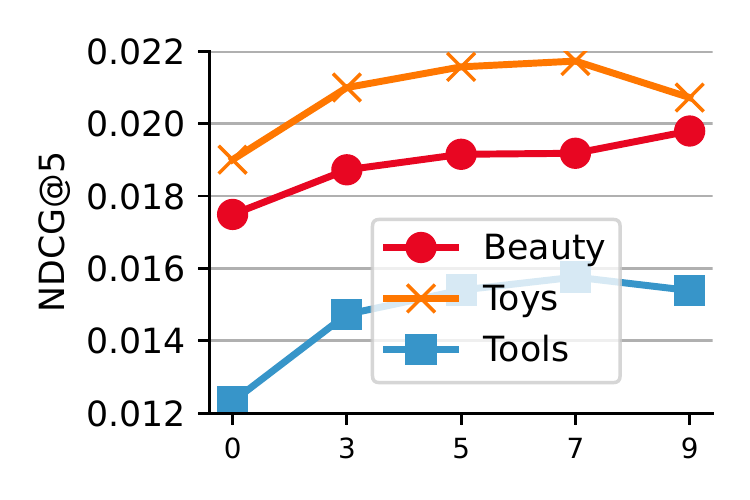}
         \caption{Different values of $U_k$ with best $I_k>0$.}
         \label{fig:Uk}
     \end{subfigure}\hfill
     \begin{subfigure}[b]{0.2\textwidth}
         \centering
         \includegraphics[width=1\textwidth]{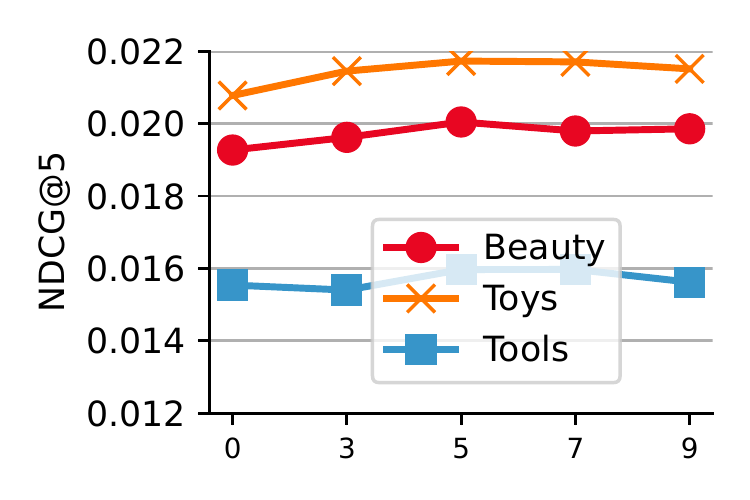}
         \caption{Different values of $I_k$ with best $U_k>0$.}
         \label{fig:Ik}
     \end{subfigure}\\
     \begin{subfigure}[b]{0.2\textwidth}
         \centering
         \includegraphics[width=1\textwidth]{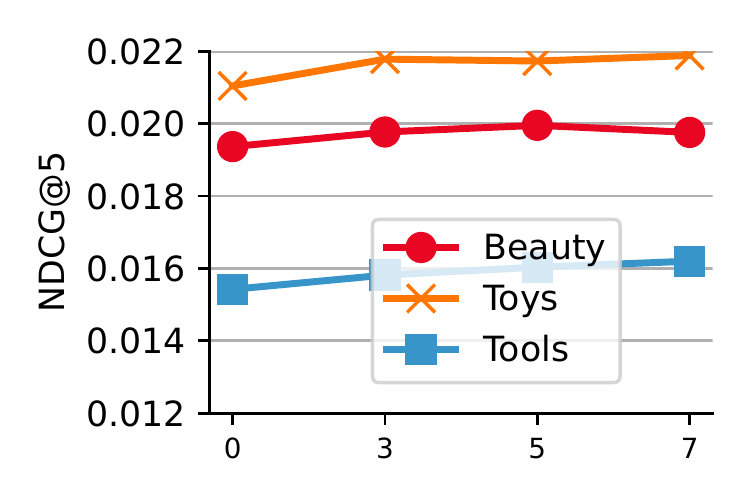}
         \caption{Different values of $UU_k$}
         \label{fig:UU_k}
     \end{subfigure}\hfill
     \begin{subfigure}[b]{0.2\textwidth}
         \centering
         \includegraphics[width=1\textwidth]{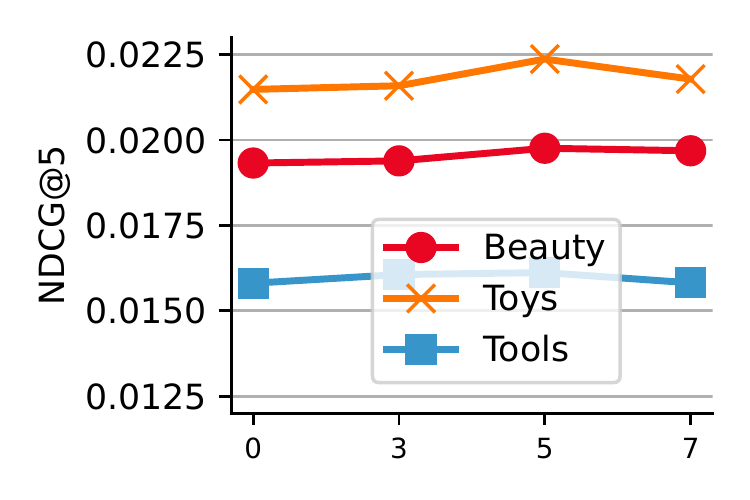}
         \caption{Different values of $II_k$}
         \label{fig:II_k}
     \end{subfigure}
    \caption{Different values of hyper-parameters, including $U_k$, $I_k$, $UU_k$ for $\mathbf{W}_{UU}$, $II_k$ for $\mathbf{W}_{II}$, with definitions in Section \ref{sec:enhanced_adj}.}
    \label{fig:perf_hyper_param}
\end{figure}

We visualize changes of NDCG@5 from hyper-parameters $U_k$, $I_k$, $UU_k$, and $II_k$, which control the number of neighbors from different semantics for purposes in Fig.~(\ref{fig:perf_hyper_param}). We have following observations:
\begin{itemize}[leftmargin=*]
    \item \textbf{The enhanced collaborative neighbors benefit the overall performance.} From Fig.~(\ref{fig:Uk}) and Fig.~(\ref{fig:Ik}), we can see that the proper choices of $U_k>0$ and $I_k>0$ improve the performances. The special cases $U_k=0$ and $I_k=0$ happen when we generate neighbors from either only the item side ($U_k=0$ and $I_k>0$) in Fig.~(\ref{fig:Uk}) or only the user side ($I_k=0$ and $U_k>0$) in Fig.~(\ref{fig:Ik}). 
    \item \textbf{The user side generated neighbors $U_k$ are important.} From Fig.~(\ref{fig:Uk}), we can observe that removing neighbors from the user side ($U_k=0$) causes significant performance degradation. From Fig.~(\ref{fig:Ik}), having more neighbors from the item side ($I_k>0$) benefits but with marginal improvements.
    \item \textbf{Small numbers of neighbors in $\textbf{W}_{UU}$ and $\textbf{W}_{II}$ are sufficient.} From Fig.~(\ref{fig:UU_k}) and Fig.~(\ref{fig:II_k}), we can observe that non-zero neighbors on $\textbf{W}_{UU}$ and $\textbf{W}_{II}$ can benefit the recommendation, which is also observed in Table~\ref{tab:overall_performances}. However, the larger values of $UU_k$ and $II_k$ do not bring significant improvements, where the lines in Fig.~(\ref{fig:UU_k}) and Fig.~(\ref{fig:II_k}) are smooth. This observation demonstrates that the potentially small $UU_k$ and $II_k$ can be sufficient.
\end{itemize}

\subsection{Improvements Analysis~(RQ3)}
\begin{figure}
    \centering
    \begin{subfigure}[b]{0.155\textwidth}
         \centering
         \includegraphics[width=1\textwidth]{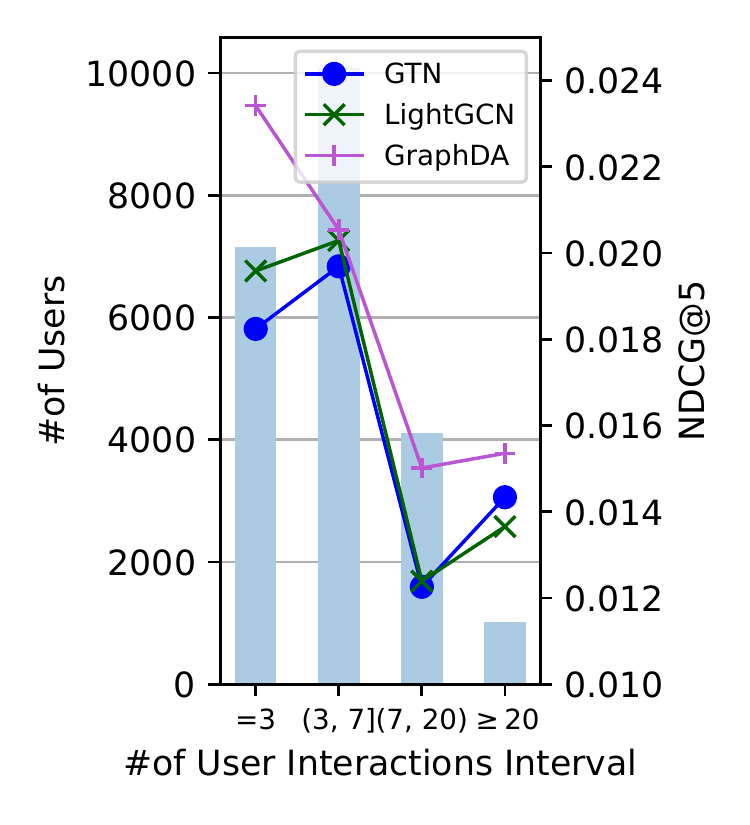}
         \caption{Beauty}
         \label{fig:beauty_users_improv}
     \end{subfigure}\hfill
     \begin{subfigure}[b]{0.155\textwidth}
         \centering
         \includegraphics[width=1\textwidth]{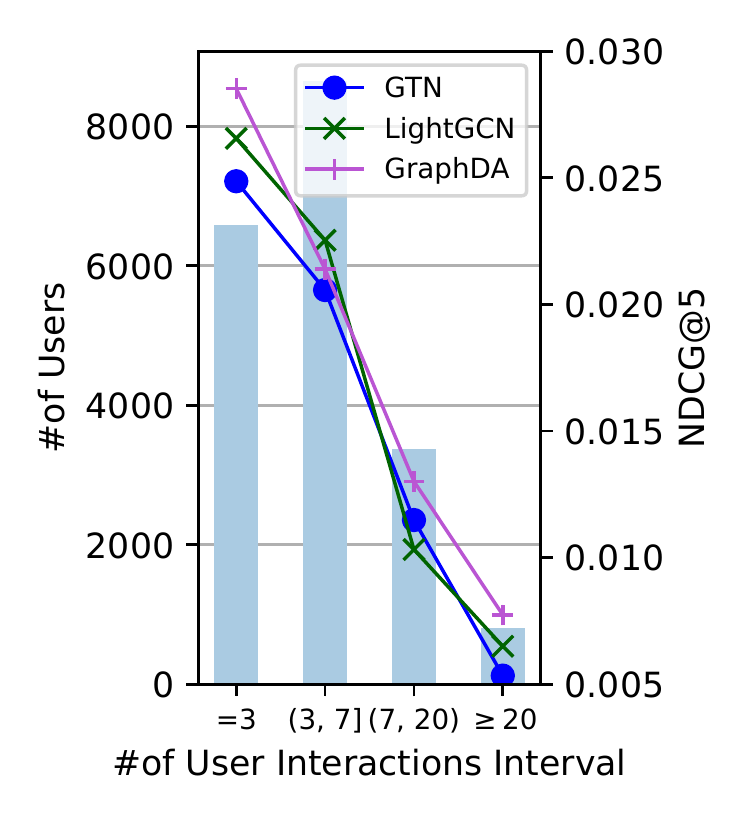}
         \caption{Toys}
         \label{fig:toys_users_improv}
     \end{subfigure}
     \hfill
     \begin{subfigure}[b]{0.155\textwidth}
         \centering
         \includegraphics[width=1\textwidth]{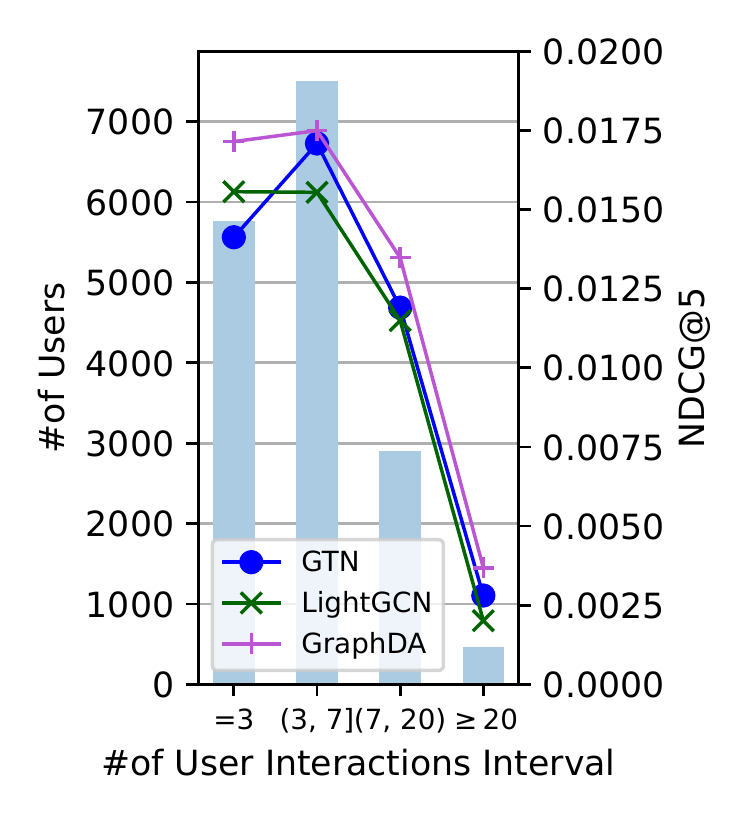}
         \caption{Tools}
         \label{fig:tools_users_improv}
     \end{subfigure}
    \caption{Improvements analysis of users grouped by the number of interactions. We only show two strong baselines to avoid cluttering figures.}
    \label{fig:perf_seqlen_dist}
\end{figure}

We validate that the proposed \modelname can denoise and augment for different groups of users, grouped by numbers of interactions, as shown in Fig.~(\ref{fig:perf_seqlen_dist}). We have following observations:
\begin{itemize}[leftmargin=*]
    \item \textbf{The augmented neighbors by \modelname benefit inactive users significantly in the graph-based recommendation.} For inactive users with least interactions (i.e., with less than 3 interactions), the proposed \modelname achieves significant improvements over the best baseline with range from 3.7\% to 19.6\%.
    \item \textbf{The denoised interactions by \modelname improve highly active users with noisy interactions, especially in datasets with potentially more noises.} For highly active users with abundant interactions (i.e., with more than 7 interactions), the proposed \modelname achieves comparative and mostly better performance than LightGCN, with improvements from 5.1\% to 67.8\%. The proposed \modelname and GTN both benefit the highly active users with a large margin over LightGCN in the Tools dataset, which potentially includes more noisy user interactions. Table~\ref{tab:overall_performances} also shows potential noises in the Tools, where the denoising GTN is the best baseline.
\end{itemize}

\section{Conclusions}
We empirically investigate the existing deficiencies of graph-based recommendations, and arguably identify that issues come from the unsatisfactory definition of the bipartite adjacency matrix. To generate a better bipartite adjacency matrix, we propose the denoising and augmentation pipeline \modelname with pre-training and enhancing steps to generate a better user-item matrix, user-user correlations, and also the item-item correlations. Experiments show the superiority of \modelname, especially for highly active users and inactive users. 
\begin{acks}
This paper was supported by the National Key R\&D Program of China through grant 2022YFB3104703, NSFC through grant 62002007, Natural Science Foundation of Beijing Municipality through grant 4222030, and S\&T Program of Hebei through grant 21340301D.
Philip S. Yu was supported by NSF under grants III-1763325, III-1909323, III-2106758, and SaTC-1930941. This work is also partially supported by NSF through grants IIS-1763365 and IIS-2106972.
\end{acks}

\bibliographystyle{ACM-Reference-Format}
\balance
\bibliography{sample-base}


\end{document}